# Parametric dynamic causal modelling

Amirhossein Jafarian*[1], Peter Zeidman[1], Rob. C Wykes[2,3], Matthew Walker[2], Karl J. Friston[1]

[1] The Wellcome Centre for Human Neuroimaging, UCL Queen Square Institute of Neurology, United Kingdom

[2] Department of Clinical & Experimental Epilepsy, UCL Queen Square Institute of Neurology, United Kingdom

[3] Nanomedicine Lab, University of Manchester, United Kingdom

***Correspondence:** Amirhossein Jafarian

The Wellcome Centre for Human Neuroimaging,

12 Queen Square, London, UK. WC1N 3AR

*E-mail:* a.jafarian@ucl.ac.uk

## Abstract

This technical note introduces parametric dynamic causal modelling, a method for inferring slow changes in biophysical parameters that control fluctuations of fast neuronal states. The application domain we have in mind is inferring slow changes in variables (e.g., extracellular ion concentrations or synaptic efficacy) that underlie phase transitions in brain activity (e.g., paroxysmal seizure activity). The scheme is efficient and yet retains a biophysical interpretation, in virtue of being based on established neural mass models that are equipped with a slow dynamic on the parameters (such as synaptic rate constants or effective connectivity). In brief, we use an adiabatic approximation to summarise fast fluctuations in hidden neuronal states (and their expression in sensors) in terms of their second order statistics; namely, their complex cross spectra. This allows one to specify and compare models of slowly changing parameters (using Bayesian model reduction) that generate a sequence of empirical cross spectra of electrophysiological recordings. Crucially, we use the slow fluctuations in the spectral power of neuronal activity as empirical priors on changes in synaptic parameters. This introduces a circular causality, in which synaptic parameters underwrite fast neuronal activity that, in turn, induces activity-dependent plasticity in synaptic parameters. In this foundational paper, we describe the underlying model, establish its face validity using simulations and provide an illustrative application to a chemoconvulsant animal model of seizure activity.

# 1. Introduction

This paper introduces a class of dynamic causal model (DCM) that can be used for characterising slow fluctuations in biophysical parameters that might underlie phase transitions in the brain. This method is based on a separation of temporal scales (Jirsa et al., 1994, Papadopoulou et al., 2017, Rosch et al., 2018a, Rosch et al., 2018b, Rosch et al., 2018c, Blenkinsop et al., 2012, Jirsa et al., 2014, Steyn-Ross and Steyn-Ross, 2010, Nevado-Holgado et al., 2012) where fast neuronal fluctuations are generated by slow fluctuations in synaptic parameters and other neurophysiological parameters (e.g., extracellular potassium). DCM then allows one to specify different hypotheses about causal relations between slow biological mechanisms (Papadopoulou et al., 2015) and select the most likely model that explains phase transitions in electrophysiological data. The innovation of the DCM introduced here is that the separation of temporal scales is used to introduce a circular causality in which synaptic parameters shape fast neuronal activity, while frequency specific neuronal activity induces plasticity or changes in synaptic parameters. In other words, fast, (spectral) neuronal dynamics are modulated on a slow timescale by drifts in synaptic parameters, while the spectral characteristics of fast neuronal activity causes slow changes in the parameters. In this paper, we illustrate how this circular causality and implicit separation of temporal scales leads to the spontaneous onset of phase transitions in brain activity and crucially, how this formulation of (patho)physiology can be used as the basis of a relatively straightforward hierarchical DCM, which we refer to as a parametric (or P)-DCM. In brief, the mapping from synaptic parameters to fast – within epoch – neuronal activity uses a conventional DCM for cross spectral density (CSD). In P-DCM, slow – between epoch – changes in spectral density are then used as an empirical prior on synaptic parameters to model activity-dependent plasticity.

The motivation for developing P-DCM was to provide people with a relatively straightforward procedure that enables them to evaluate hypotheses about the underlying causes of phase transitions in neuronal activity, in terms of model evidence. For instance, the circular causality between slow parameters and spectral responses of neuronal oscillations could be useful for understanding the relationship between depth of anaesthesia (induced with gradual drug injections such as Propofol to modulate the frequency contents of electrophysiological recordings) and slow dynamics of synaptic efficacy of neuronal populations (Purdon et al., 2013, Hashemi and Hutt, 2016). Parametric DCM could be useful for investigating how the slow evolution of ion/synaptic mechanisms engenders brain state transitions, e.g., awake to asleep (Muheim et al., 2019) or pathological states, e.g. seizures (Grenier et al., 2003). Exemplar questions we envisage being addressed using P-DCM include: (i) what sorts of slow biological mechanisms might account for the decline of alpha band power as a marker of progression in Alzheimer's disease? (Li et al., 2020); (ii) what are the links between induced paroxysmal epileptic transitions (3 to 8 Hz) and pathological evolution of potassium level or synaptic transmission mechanisms? (Moody Jr et al., 1974, Du et al., 2016); (iii) what is the role of changes in synaptic efficacy in inhibitory populations and the emergence of high beta activity, as hallmarks of pathophysiology in Parkinson's disease (McCarthy et al., 2011)? We do not address these long-standing questions here. Instead, through a worked example of P-DCM we show how one can (i) establish causal links between biological parameters and spectral responses in generative models of neuroimaging data; (ii) evaluate model evidence and infer parameters from empirical data and; (iii) compare different hypotheses (through Bayesian model reduction and comparison) about how the data were generated.

With regards to extant procedures, P-DCM complements previous approaches to fast-slow modelling of phase transitions (Steyn-Ross and Steyn-Ross, 2010, Coombes and Bressloff, 2005, Baier et al., 2012, Wendling et al., 2016, Schiff, 2012). Conventionally, multiscale models feature fast states that constitute neuronal dynamics, which are (predominantly) regulated by the dynamics of some slow states, such as ion concentrations and synaptic efficacy. The prominent application of multiscale models is to understand the taxonomy and phenomenology of phase transitions in the neuroimaging data, in particular electrophysiological recordings, as reviewed in (Friston, 2014). Some recent (selective) examples of these models are: (i) a neuronal-glial interaction model that emulates epileptic seizures and cortical spreading depression through slow dynamics of potassium and sodium (Wei et al., 2014); (ii) a neural mass model (NMM) augmented with slow dynamics for synaptic efficacy—e.g. as a model of synaptic plasticity (Fung and Robinson, 2014)—to replicate burst suppression in anaesthesia (Liley and Walsh, 2013, Liu and Ching, 2017); (iii) a slow-fast mesoscale model of epileptic seizures that captures the slow dynamics of firing thresholds—to replicate spike rate adaptation—with a NMM (Jafarian, 2016, Jafarian et al., 2019b); and (iv) a phenomenological model of epileptic seizures, known as an Epileptor, where the aetiology of epileptic seizures is explained via the evolution of a slow state (Jirsa et al., 2014, El Houssaini et al., 2020). The novelty of parametric DCM, as a forward model, is the ability to link slow biological mechanisms with the spectral contents of fast neuronal dynamics. In short, P-DCM was designed to characterise (paroxysmal) transitions formally, by coupling modulatory slow dynamics of ion current/synaptic efficacy and induced spectral responses in mesoscale dynamics.

In terms of model estimation, due to the complexities of multiscale models, inferring the parameters of these models—and comparing models given real electrophysiological recordings—typically incurs a high computational burden (Hashemi et al., 2020, Jafarian et al., 2019b, Jafarian et al., 2019a). In this paper, we show that P-DCM provides a computationally efficient method for parameter estimation—and more importantly calculating model evidence—that can be used to investigate the aetiology of phase transitions from electrophysiological recordings. In particular, model inversion in P-DCM is motivated by Synergetic theory and the Adiabatic approximation (Basar et al., 2012, Haken, 1977, Jirsa and Haken, 1997). This formulation assumes that for each value of slowly varying parameters, the fast-neuronal states attain nonequilibrium steady-state that is expressed in terms of spectral contents. This assumption leads to a relativity straightforward approach to inferring model parameters from empirical data. Furthermore, P-DCM enables one to compare different models or hypotheses about underlying generators of paroxysmal transitions.

This paper comprises four sections. In the next section, we review the theoretical tenets of P-DCM. In section three, we first provide an illustrative example of the basic ideas in terms of a forward simulation of the generative model. This example illustrates how changes in model parameters can induce transitions in brain activity and accompanying spectral responses. We then use these forward simulations to create a hierarchical generative model (Friston et al., 2016) that can be inferred from empirical data. The third section presents some applications of P-DCM—using Bayesian model reduction—to characterise the underlying causes of seizures. In this section, we generated synthetic data that undergoes transition into and out of seizures. Then, by performing a second level analysis, under a hierarchical model of slow changes in DCM parameters (Friston et al., 2015), we show how one can answer some fundamental questions concerning the genesis of epileptic seizures. This section can be read as establishing the face validity of the procedure. In the fourth section, we apply model inversion to an empirical electrophysiological recordings from an animal model of epilepsy to provide an illustrative (worked) example of this type of analysis. We conclude with a discussion of the limitations and potential applications of parametric DCM.

# 2. Theory

This section provides a brief review of dynamic causal modelling. Then, we build on this to introduce the P-DCM methodology that uses a hierarchical model, in which slow changes in DCM parameters at the second level are coupled to spectral responses at the first.

## 2.1 Dynamic Causal Modelling

Dynamical causal modelling is the estimation of biologically informed models of neuroimaging data using variational Bayesian methods (Friston et al., 2007, Friston et al., 2003, Friston et al., 2008, Friston et al., 2019). DCM was pioneered nearly 20 years ago and since been used to infer the biological mechanisms generating neuroimaging data (Jafarian et al., 2019c, Friston, 2011, Friston et al., 2011, Penny et al., 2011, van Wijk et al., 2018). In DCM, a posterior probability density over model parameters, as well as the evidence for a model (for any given empirical data), are inferred through optimisation of an objective function called the variational free energy. This objective is known as an evidence lower bound (ELBO) in machine learning and provides a computationally efficient approximation to the (log) model evidence or marginal likelihood. This optimisation is performed under the Laplace approximation (i.e., probability densities are approximated using Gaussian distributions) (Friston et al., 2007, Friston et al., 2008, Beal, 2003, Zeidman et al., 2019a) using a gradient ascent on variational free energy. This is known as Variational Laplace. In DCM, the model evidence associated with different hypotheses (i.e., models) of the same data are compared using Bayesian model selection and comparison to identify the best explanation for the data at hand (Kass and Raftery, 1995). A recently developed Bayesian model reduction technique, which we will leverage here, opens a new avenue for rapidly comparing the evidence for models specified in terms of their priors (Friston et al., 2018, Friston and Penny, 2011, Friston et al., 2016, Zeidman et al., 2019b).

## 2.2 Parametric dynamic causal model (P-DCM)

2.2.1 Theoretical foundation

The starting point for P-DCM is a separation of temporal scales into fast (neuronal) and slow (parametric) fluctuations (Basar et al., 2012). Our main assumption here is that the brain can be treated as an excitable medium, where neuronal dynamics operate around a sequence of fixed points, where oscillatory dynamics result from the (neuronal) filtering of endogenous random fluctuations (Moran et al., 2007, Moran et al., 2013, Friston et al., 2012). The fixed points then change slowly to generate changes in cross spectral density over time.

The slow and fast scale separation, which is known as an adiabatic approximation (Haken, 1977), leads naturally to a mean field approximation when building generative models for data analysis. Under a mean field approximation, the posterior over unknown variables is approximated with the product of (marginal) posteriors (Friston et al., 2007). In this instance, the adiabatic approximation enables us to estimate slowly changing synaptic (i.e. connectivity) parameters and hyperparameters, under the assumption that they are conditionally independent. The mean field approximation requires us to write down the equations of motion for the parameters or connection strengths. Effectively, this entails specifying a model of synaptic plasticity. We therefore consider a paroxysmal (e.g., epileptogenic) perturbation or phase transition that is mediated by neuronal plasticity (e.g., spike rate adaptation or associative plasticity) to reset the fast-neuronal dynamics – and their expression in spectral responses.

In P-DCM, neuronal dynamics are generated in terms of their second order (statistical) moments, via complex cross spectra (Friston et al., 2012). These then generate observed cross spectra in the sensor domain (via a standard electromagnetic forward model). In detail, let us write the temporal dynamics of neuronal states, $x$, driven by random fluctuations, $u$, for a given set of parameters $\theta$ as follows:

$$\dot{x} = f(x, \theta) + u \qquad (1)$$

In equation 1, the cross spectral content of the random fluctuation, $g_u(\omega, \theta) = FT(E[u(t), u(t - \tau)])$, is modelled as (structured) pink noise (Friston et al., 2012, Moran et al., 2007, Moran et al., 2013). The assumption underlying DCM for CSD is that random fluctuations induce oscillations around some fixed point (Da Silva et al., 1974, Friston et al., 2012). In other words, using DCM for CSD, we treat electrophysiological recording, $y$, as a neuronally filtered version of endogenous noise (Da Silva et al., 1974). Under a fixed point assumption, the neuronal dynamics can be well approximated by the first order linearised neuronal model $\dot{x} = (\nabla_{x^*}f)x + u$ (where $\nabla_{x^*}$ denotes the Jacobian of neuronal states at $x^*$) (David et al., 2006). The linearised equation of neuronal states is used to obtain a semi analytic expression of the transfer function (i.e., first order Volterra kernel) with an impulse response $k(\tau, \theta) = \exp(\tau. \nabla_x f(x, \theta))$. The spectral response of the transfer function can in turn be expressed as follows:

$$K(\omega, \theta) = FT\ (\exp \tau. \nabla_x f(x, \theta)) \quad (2)$$

The neuronal source response, $g_x(\omega)$, of the noise driven model is given by:

$$g_x(\omega) = K(\omega, \theta). g_u(\omega, \theta). K(\omega, \theta)^T + g_o(\omega, \theta) \quad (3)$$

In equation 3, $g_o(\omega, \theta)$ represents random fluctuations due to observation noise (associated with individual channels and common to all channels) (Moran et al., 2007). The spectral response in sensor space is calculated through the forward electromagnetic model, denoted by $L.M(\omega)$, as follows:

$$g_y(\omega) = L.M(\omega). g_x(\omega). M^T(\omega). L^T + \epsilon \quad (4)$$

In equation 4, $g_y(\omega)$ are the cross spectral data and $\epsilon$ is a random effect due to computing the cross spectra from finite timeseries, which can be inferred using the variational Bayes approach in DCM (please see appendix A for explanation of this final term in equation 4).

In P-DCM, slow dynamics of parameters, $\theta$, have the following general form, which models activity-dependent plasticity, via a dependency on neuronal activity:

$$\theta_t = \mathfrak{S}_\rho\big(g(\omega, t)\big) + \epsilon_t \quad (5)$$

The term $g(\omega, t)$ in equation 5 denotes the spectral activity of neuronal dynamics at epoch $t$. Here, $\mathfrak{S}$ is an operator (parametrised by unknown parameters $\rho$) that generates the parameter at time $t$ from the spectral activity of neuronal dynamics, and $\epsilon_t$ is additive random effect. One example of the operator $\mathfrak{S}$ could be the integral of power spectral density (PSD) (the integral of the PSD is equal to the variance of signal in time domain), which engenders a simple form of synaptic plasticity. Another example of operator $\mathfrak{S}$ can be concatenated power spectral density of electrophysiological data, e.g., local field potential (LFP) over different epochs. Such definitions for the $\mathfrak{S}$ operator can be useful when one wants to test for a causal relation between synaptic parameters and frequency specific neuronal activity (i.e., investigating circular causality). In what follows, we will use a general linear model (GLM) based upon regressors (in a design matrix) that encode the expression of particular frequencies. This means that the parameters $\rho$ correspond to the parameters of a GLM (e.g., regression coefficients). Note that the adiabatic approximation allows us to express the slow dynamics as a mapping from the time-dependent spectral density to the parameters.

In P-DCM, the neural mass model is only used to evaluate the spectral response to some endogenous neuronal fluctuations $u(t)$ that are themselves, parameterised. This enables us to specify a generative model of slow fluctuations in observable (complex cross) spectral density, purely in terms of parametric dynamics, while – at the same time – absorbing a chosen neural mass model into the resulting parametric DCM.

2.2.2 Practical Implementation

For people familiar with DCM, we employ a standard DCM for cross spectral density (Litvak et al., 2011, Friston et al., 2012) and turn it into a hierarchical state space model by adding dynamics to the parameters. The inversion of this model allows us to estimate the connectivity that best explains empirical cross spectra and the trajectories or dynamical architecture controlling the expression of different synaptic parameters.

Practically, we first divide data into epochs (that may overlap). We then use the separation of temporal scales to model slow changes in the parameters from epoch to epoch within a hierarchical or parametric empirical Bayes model (Friston et al., 2015, Friston et al., 2016). In detail, the posteriors over parameters from each epoch are passed to the second level, and slow drifts in the parameters are captured using a general linear model (GLM) (see Papadopoulou et al., 2015). Crucially, the explanatory variables in the GLM comprise the spectral density of neuronal activity in each epoch. This underwrites the circular causality between frequency specific changes in fast, parameter-dependent neuronal responses and the slow, activity-dependent neuronal plasticity. In the examples below, we use the empirical spectra as explanatory variables in the GLM – after some suitable dimension reduction and transformation (Chen et al., 2008, Gavish and Donoho, 2014). This means the empirical spectra are used twice. First, the cross spectrum of each epoch is used as the data feature to fit the parameters of an epoch-specific DCM. Second, between-epoch changes in spectral activity are used to provide empirical priors over changes in these parameters. This dual use of the spectral data is licensed by the separation of temporal scales upon which this DCM rests. In other words, P-DCM leverages information in the spectral content of each epoch and in spectral changes over epochs.

The reason that this (parametric) DCM is efficient is that using a generative model of cross spectra converts a neural state-space model into an instantaneous mapping between the parameters of the model and the expected second order responses over time; namely, the complex cross spectra. The only assumption behind this adiabatic approximation is that the spectral summary of dynamics is sufficient to inform slow changes in parameters. This is a key part of the parametric DCM described here: conditioning the parameters of neuronal density dynamics (here, spectral density) means that we can model activity-dependent changes in connectivity and other slowly varying factors that, in turn, shape fast neuronal responses. This means that one can specify a parametric DCM to model slow dynamics such as spike rate adaptation, short-term plasticity or, indeed, the target of this work; trajectories in parameter space that engender paroxysmal transitions in neuronal dynamics, e.g., epilepsy. To establish the face validity of this kind of model, we will refer to specific empirical data in which seizures were induced chemically. These data and the ensuing seizure activity and now summarised briefly.

2.3 Chemoconvulsant animal model of seizures

Animal experiments were conducted in accordance with the United Kingdom Animal (Scientific Procedures) Act 1986, and approved by the Home Office (license PPL70-13691).

Sprague-Dawely rats (8–12 weeks old, 280–330 g; Charles River, UK) were used in this study. All animal experiments were conducted in accordance with the United Kingdom Animal (Scientific Procedures) Act 1986, and approved by the local ethics committee (University College London). Animals were housed on 12 h/12 h dark/light cycle, and food and water were given ad libitum. Animals were group housed and allowed to acclimatise to the new environment for at least 1 week before surgery, and were housed individually after surgery. Rats were anaesthetised using isoflurane (2%) and head-fixed in a stereotaxic frame (Kopf, USA). A small hole was drilled through the skull above the right primary visual cortex (coordinates: 3 mm lateral and 7 mm posterior of bregma and a cannula inserted (Plastics1, USA). During the same surgery, an ECoG transmitter [A3028E-AA, Open Source Instruments] was implanted subcutaneously with a recording electrode wire positioned in the visual cortex. A reference electrode was placed in the contralateral frontoparietal cortex. Animals were single housed in Faraday cages for telemetric ECoG recordings. 7-10 days post-surgery, rats were briefly anaesthetised and 300-400nl of 10mM Picrotoxin administered to layer 5 visual cortex, via the pre-implanted cannula. Immediately after injection rats were removed from the stereotaxic frame and replaced in telemetry. Within a few minutes post-injection small amplitude spikes appeared in the ECoG traces, these evolved over the next 5-10 minutes into large amplitude regular (~1Hz) spikes or poly-spikes. A cyclical pattern of seizures (~53s in duration) and inter-ictal activity was observed for about 2-3 hours before fading away and a resumption of normal brain activity.

## 3. Validation analyses using simulated data

This paper demonstrates the basic phenomenology that P-DCM is capable of explaining; namely, phase transitions in electrophysiological data. In this section, we will use synthetic data. In subsequent sections, we apply the same methodology to empirical data taken from the mouse model of seizures above, to illustrate the sort of analyses one can perform.

To illustrate P-DCM, we use the canonical microcircuit (CMC) (Bastos et al., 2012, Friston et al., 2019) as a model of electrical activity of a typical cortical column (see Figure 1 for details), This model has been found to be useful for a range of applications in computational neuroscience, including: predictive coding (Bastos et al., 2012), modelling evoked brain responses (Auksztulewicz and Friston, 2015, Jafarian et al., 2020, Friston et al., 2019) and cross spectral responses (Rosch et al., 2018b, Rosch et al., 2018c) to name a few. The CMC can replicate fast gamma activity of the superficial layers as well as the slow (beta) activity in deep layers in the cortex (Bastos et al., 2012). The CMC comprises four neuronal populations, namely superficial pyramidal cells (layer I of the cortex), excitatory populations (spiny cells in layer IV of the cortex), deep pyramidal cells (layer V of the cortex) and interneuron inhibitory cells. Each population in the CMC receives a firing rate from inter regional and distal neuronal populations, weighted by intrinsic and extrinsic effective connectively, respectively. Each population converts summed input firing rates to a postsynaptic synaptic response (through convolution of the firing rate and a synaptic impulse response model). The ensuing postsynaptic response generated by each neuronal population is then transformed to mean firing rates (through sigmoid transformation), which is then communicated to other populations via intrinsic and extrinsic efferents. Electrophysiological recordings are generated by the CMC as (mainly) the activity of superficial pyramidal cells plus weighted sum (inferred from data) of the activity of excitatory and deep pyramidal populations.

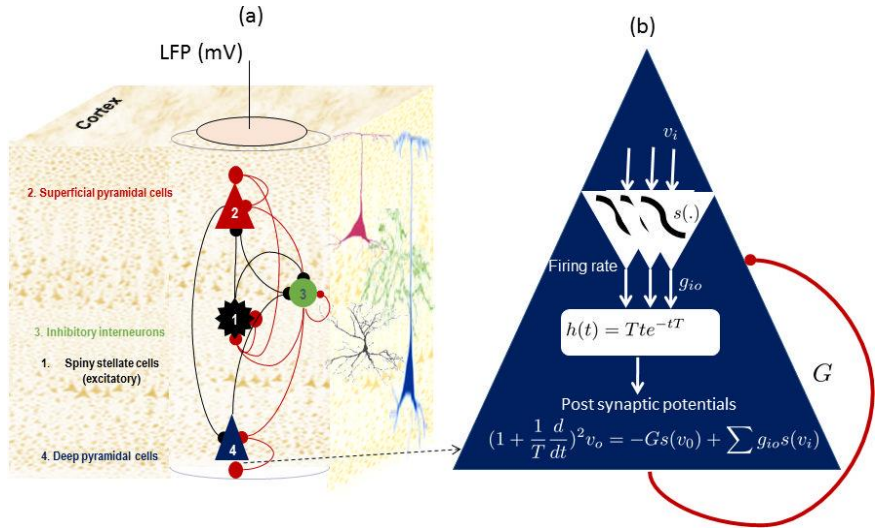

*Figure 1. Canonical microcircuit (CMC). a) A patch of cortex divided into cortical columns. Electrical activity of each cortical column can be captured by electrophysiological recording, e.g., ECoG. Each cortical column is divided into several layers (here 3), each of which is modelled by one population of neurons. Superficial and deep pyramidal cells are in the superficial and deep layers, respectively. Excitatory interneurons (spiny stellate cells) are located in layer four – labelled 1 in the figure. Inhibitory interneurons are distributed across all layers and are modelled using one population that interacts with all other populations. b) The mean electrical activity of each neuronal population is derived using mean field theory using two conversion operators. The postsynaptic potentials, $x_i$ are transformed through a sigmoid nonlinearity, $s(.)$, To generate a firing rate (weighted by connectivity constant g). The ensuing firing rate is then converted to postsynaptic potentials, $x_o$ through the linear response of synapses (parametrised by rate constant T). In addition, each population is equipped with a self-*

*inhibition connection G (illustrated as short curved red lines) which assures neuronal homeostasis, i.e., in the absence of neuronal input, the activity of neurons rest at an equilibrium.*

## 3.1 Face validity: simulation of beta burst synchronization

Part A: Forward simulation of beta bursts

In this section, we validate the basic idea of P-DCM – that a reciprocal coupling between fast neuronal activity and slow drifts in synaptic parameters (e.g., extracellular concentrations or synaptic efficacy) induces phase transitions. This validation rests upon integrating (or solving) coupled differential equations at fast and slow timescales to illustrate that when parameters pass from one regime of parameter space to another, there is a qualitative change in the spectral activity at the fast timescale (Papadopoulou et al., 2015).

In this simulation, we illustrate the effect of modulating the self-inhibition of deep and inhibitory cells using the integral of the power spectral density (variance of the signal in the time domain) of their postsynaptic potentials. This can be regarded as a simple model of activity-dependent synaptic plasticity (Fung and Robinson, 2014, Fritschy, 2008). This model was motivated by the fact that there is a relationship between the energy content of neuronal activity and ion dynamics. For example, the activity level of neuronal activity declines after seizures (known as post ictal depression), which is related to the pathological evolution of ion currents (Panayiotopoulos, 2010). Another motivation includes the relationship between energy metabolism due to neuronal activity (caused by ion dynamics) and their level of demand for energy as observed in haemodynamic responses (Rosa et al., 2011, Carmichael et al., 2017).

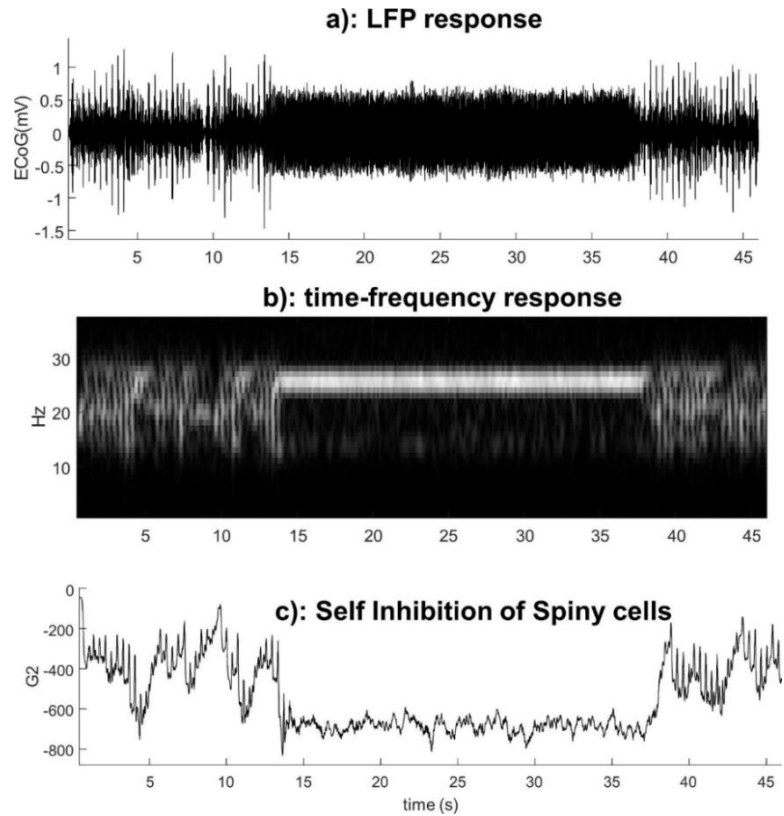

*Figure 2. Forward simulation of the CMC model with slow dynamics. a) LFP response of the model where self-inhibition of spiny and inhibitory populations showed slow changes that depend on postsynaptic potentials generated by spiny and inhibitory cells, respectively. Clearly, there are two distinct dynamics can be identified in the time domain - spiky normal activity and synchronised activity in the middle. b) Time frequency representation shows the frequency of synchronised activity in the beta range. c) Slow fluctuation of parameters before and after phase transitions- in this case there is a clear separation between the value of the parameter that produced the two types of activities in LFP. In part c, $G_2$ is the self-inhibition of excitatory population.*

The ensuing simulated data and evolution of parameters are shown in Figure 2. In this simulation, the dynamics of parameters induce a high band synchronised beta burst. The dynamics of phase transition in this model are akin to a bifurcation or phase-transition. This is because crossing a phase boundary or 'separatrix' in parameter space induces the transitions in mesoscopic activity. Although there is a sudden change in the spectral activity induced by this boundary crossing, the drift of the parameters *per se* is quite smooth and slow (in comparison with the fast neuronal states) (Papadopoulou et al., 2017, Rosch et al., 2018a, Rosch et al., 2018c). The synchronised beta burst is a hallmark of movement disorders (McCarthy et al., 2011, Spitzer and Haegens, 2017, Sherman et al., 2016), which may also be induced by drugs/interventions (Rodriguez et al., 2004, Shin et al., 2017) or during memory retrieval (Jansen et al., 2011). Note that the simulated seizure activity in Figure 2 is entirely self-organised. In other words, the drifts in self-inhibition were driven by activity-dependent plasticity; here, the overall power in neuronal activity of the respective neuronal populations.

The example in Figure 2 was based upon solving differential equations for fast and slow dynamics in time. To examine the equivalent characterisation in frequency space, we evaluated the transfer function of the CMC model, given the trajectory of parameters in Figure 2. In detail, we took the trajectory of parameters from Figure 2-c, and calculated the induced spectral changes through simulation of the transfer function for each sample point in the data. This first-order approximation to the nonlinear solution in figure 2 confirms that phase transitions simulated above can be attributed to the evolution of parameters. Furthermore, it licenses the use of spectral data features for inversion of a DCM for cross spectral density (CST) data. We pursue model inversion in the next section.

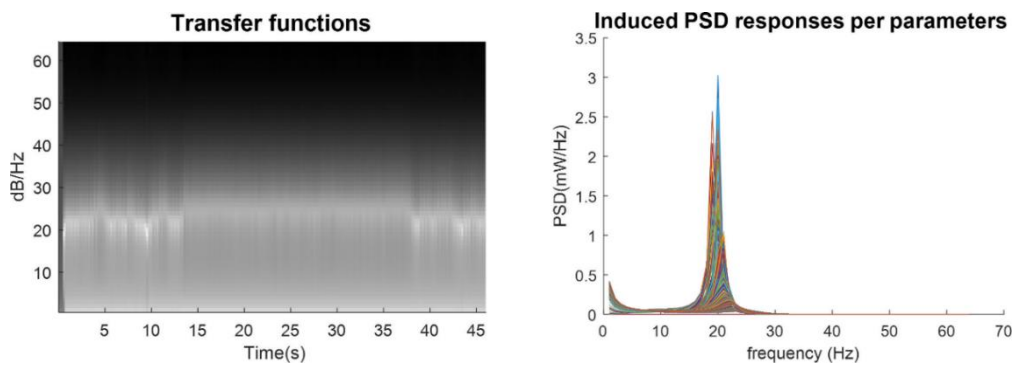

*Figure 3. Forward simulation of the transfer function CMC with slow dynamics. The figure on the left-hand side shows the time frequency response of the transfer function given the trajectory of slow parameters in figure 2 panel c. The right-hand side of the figure shows the power spectral density of the simulated transfer function for each value of slowly varying parameters in figure 2 panel c (mapped into the time frequency domain on the left figure).*

Part B: Inferring parametric dynamics

The previous section established the face validity of the generative model in terms of being able to generate plausible phase transitions. In this section, we turn to the face validation of inference or model inversion. In brief, we now try to recover the slow changes in synaptic parameters given the (synthetic) electrophysiological data in the previous section. To infer dynamics of the parameters, first we divide the simulated data above into overlapped epochs (sliding windows that span all samples in the data with the duration two seconds) . Then, we implement P-DCM by using the standard (canonical microcircuit) DCM for CSD to explain the cross spectral density of each epoch, furnishing a set of parameters for each window (Moran et al., 2007, Bastos et al., 2015). The ensuing predictions of spectral responses is shown in Figure 4.

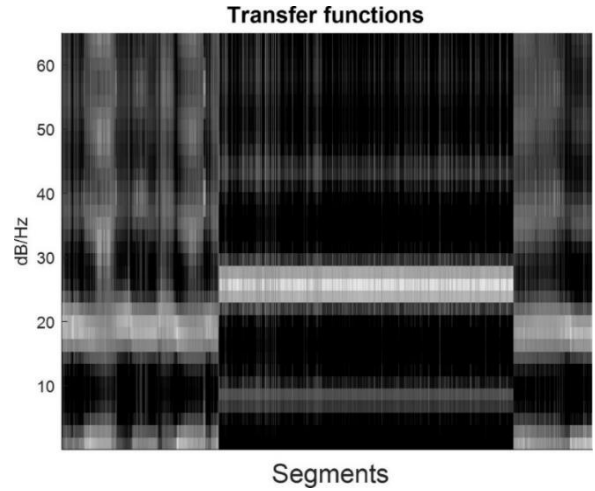

*Figure 4. Predicted time-frequency response of the estimated CMC over epochs. DCM for CSD is used to estimate parameters for each epoch. The ensuing estimates are then used to estimate the transfer functions of the CMC model, to generate predicted induced frequency responses.*

Next, to leverage the circular causality between slow and fast timescales, we inverted a general linear model (GLM) of the parameters from each epoch. The GLM we use for this example is a polynomial expansion of the spectral power:

$$\theta = \beta_1 (g(y)^1) + \beta_2 (g(y)^2) + \cdots + \beta_k (g(y)^p) + \epsilon \qquad (6)$$

In equation 6, $\beta_i$s are unknown coefficients in the GLM, $g(y)^i$ is the restricted power spectral density (to some frequency bins) of the observation data, where $i$ is an exponent that can range from one to an unknown value $p$. The frequency bins of interest, here, are dominant spectral peaks (which can be identified automatically via a singular value decomposition) before, during and after phase transitions in the observed signal. To define an optimal value for $p$, we gradually increased it from one, and compared the associated free energy (i.e., log evidence). At some point during the increase of the polynomial order, the free energy (which is the accuracy minus the complexity of the model) starts to decrease due to overfitting (Bishop, 2006). The implicit Bayesian model comparison enables one to find the polynomial order $p$ that has the greatest evidence, i.e., the best balance between accuracy and complexity. The results of this model comparison are shown in Figure 5.

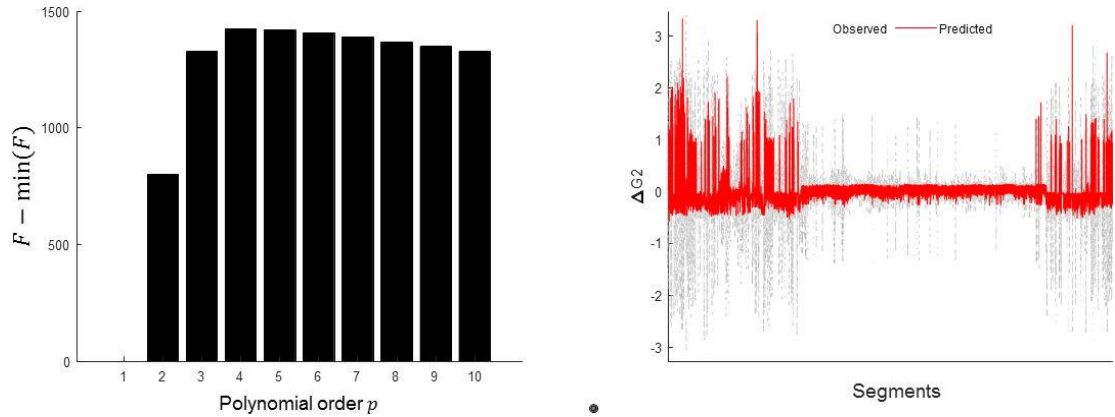

*Figure 5. Modelling the evolution of parameter ($\Delta G_2$) with respect to empirical frequency content in the data. a) This panel shows relative free energy (with respect to the smallest model evidence) of the second level GLM. We estimated the model evidence associated with models of changes in synaptic parameters based on a linear-in-coefficients polynomial of the empirical frequency content of the ensuing neuronal responses – as captured by the data (equation 6). Here the optimal power of the polynomial is 4. b) Observed and predicted model of the parameter dynamic ($\Delta G_2$) with polynomial order 4 in equation 6. In this and all subsequent figures, changes in the parameters are expressed in terms of log scaling. In other words, a value of 0 corresponds to a scaling by exp(0)=1. For small changes, the log scaling is roughly equal to the proportional change (i.e., -.1 is roughly a decrease of 10%).*

Note that by construction, the spectral data features are both caused by the parameters (at the first or fast level) and also cause the parameters (at the slow or second level). The advantage of this unusual but straightforward construction is that one can test the hypothesis that particular frequencies are responsible for increasing or decreasing particular parameters on a slow timescale. It is this influence of frequency-specific modulation on synaptic parameters that stands in for the reciprocal coupling between fast and slow timescales evinced in the first section.

3.1 Face validity: model comparison

In this section, we present a face validation of the Bayesian model comparison and reduction procedures in P-DCM that can be used to test different hypotheses about how slow parametric (synaptic) changes give rise to spontaneous paroxysmal transitions. To generate synthetic data, we specified trajectories of the self-inhibition of inhibitory and deep pyramidal cells, to produce a characteristic change in spectral density in electrophysiological data (that stands in for the onset of seizure activity). The timeseries solution of the stochastic difference/differential equations generating parameters and simulated data are shown in Figure 6. In this simulation, pathological transition is due to changes in self-inhibition of the inhibitory and deep pyramidal populations. Interestingly, we observe hysteresis phenomena in the behaviour of the model. This is supported by the fact that when self-inhibition of deep pyramidal and inhibitory cells are increased/decreased, respectively, the model undergoes high frequency activity, which is known to be one of the hallmarks of onsets of spontaneous seizures (Traub et al., 2001, Grasse et al., 2013). Importantly, when the parameters change in the other direction, the model generates epileptic spikes.

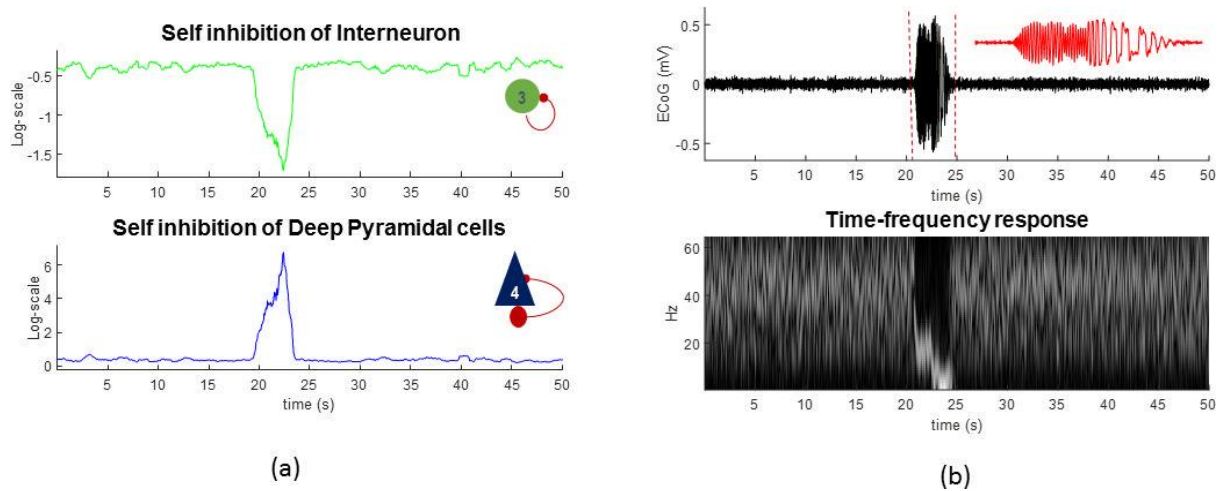

*Figure 6. Simulation of induced seizures. a) Slow dynamics of parameters of the CMC model. b) Simulated ECoG and its time frequency representations, which shows some features of spontaneous seizures, i.e., before the seizures we observe some sorts of fast activity which are followed by pathological spikes around 5-8 Hz.*

Next, to infer the parameters over 21 predefined epochs, we used DCM for CSD. The reader should note that due to the nonlinear nature of the model, there are many combinations of parameters (e.g. synaptic time constants, connectivity parameters) that could provide an equally good fit to the data (Jansen and Rit, 1995) and some may have even better free energy scores than the model generating the data (Litvak et al., 2019, Friston et al., 2013). This may sound counterintuitive; however, recall that the free energy is the trade-off between the accuracy and complexity of the model, where complexity is the difference (KL-divergence) between the priors and posteriors. Slight changes from the prior expectations in a large number of parameters may offer a less complex explanation for the data than having just a few parameters with a large deviation from their priors. Therefore, a higher evidence may be afforded to a model that is simpler than the one used to generate the data.

In this example, we first define a prior over all model parameters and then test various models in which one parameter is allowed to fluctuate around its prior expectation over epochs. To identify the prior expectation, we estimated all model parameters from a normal segment of the data in Figure 6 (alternatively, one could use several normal segments and use Bayesian model averaging to define the requisite expectation).

After identifying the prior expectation for the parameters, we ran DCM for CSD for each epoch to quantify changes in self-inhibition associated with each population. In total there are four self-inhibition connections. In other words, the hypothesis that we evaluate here is that pathological disinhibition in neuronal populations induces paroxysmal transitions and our key question of interest is: which populations contribute to the genesis of seizures? Answering such a question could have a substantive impact on our understanding of epileptic seizures (as opposed to non-epileptic seizures), which in turn may assist in designing an effective treatment strategy to supress or abate pathology.

Practically, we took the above parameter estimates (posterior mean and covariance) to a second level parametric empirical Bayes (PEB) analysis (Friston et al., 2016), with a between-epoch design matrix containing empirical priors based on the (binarized) spectral envelope of the data. These empirical priors tell us *when* particular synaptic parameters change; enabling model comparison to identify *which* particular combinations of parameters best explain the data. As it shown in Figure 7b, Bayesian model reduction suggests that changes in self-inhibition in inhibitory and deep pyramidal cells are the best explanation for the data. This is consistent with how the data were actually generated. Bayesian model reduction is used to eliminate redundant parameters by testing the evidence for models with and without a particular parameter and computing the Bayesian model average.

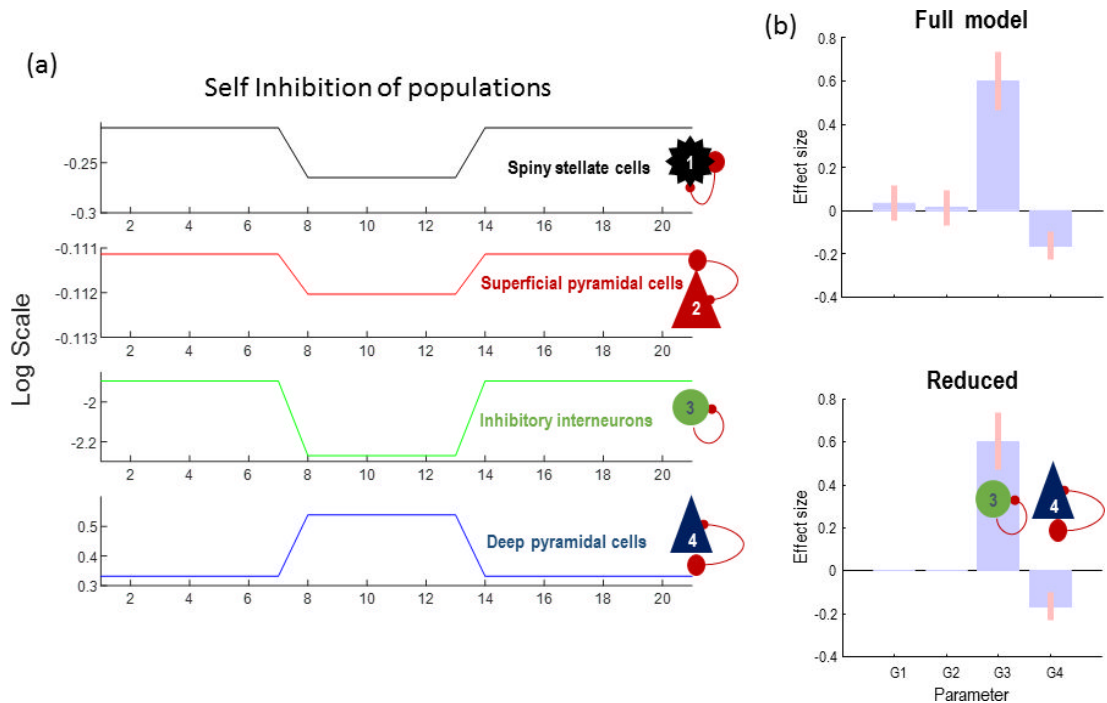

*Figure 7. Parametric empirical Bayes analysis of simulated data. a) Estimated slow dynamics of parameters in the CMC mode over 21 segments of data with average of 2 second duration. b) PEB analysis with a design matrix based on the spectral envelope of the data. The original PEB posterior estimates show on top (full model) and the estimates following Bayesian model reduction are shown below (reduced). The reduced model reveals the underlying causes of seizures in figure 6 and clearly capture is the trends in parameters. $G_1$, $G_2$, $G_3$ and $G_4$ are self-inhibition of excitatory, superficial pyramidal, inhibitory, and deep pyramidal cells, respectively. As in previous figures, the effect sizes expressed in terms of a log scaling.*

*We then repeated the PEB analysis by replacing the binarized spectral envelope with frequency specific regressors, to characterise the relationship between the self-inhibition and spectral responses. This kind of analysis could disambiguate the contribution of distinct neuronal populations to paroxysmal brain activity (which can be targeted by different intervention mechanisms). We selected three frequencies as the regressors for the PEB design matrix (i.e., GLM). Specifically, we identified the two frequencies that predominated during the seizure period and normal activity. Then used PEB to characterise the contribution of parametric changes to spectral responses in the data (or vice versa). The result of this analysis is shown in*

figure 8, and suggests that self-inhibition of the inhibitory population is likely to be responsible for the generation of 3 Hz oscillation (or vice versa) and both inhibitory and deep pyramidal self-connections are implicated in the generation of 8 Hz activity. Figure 8 shows the posterior estimates of the second level (GLM) parameters (on top) and following Bayesian model reduction (below).

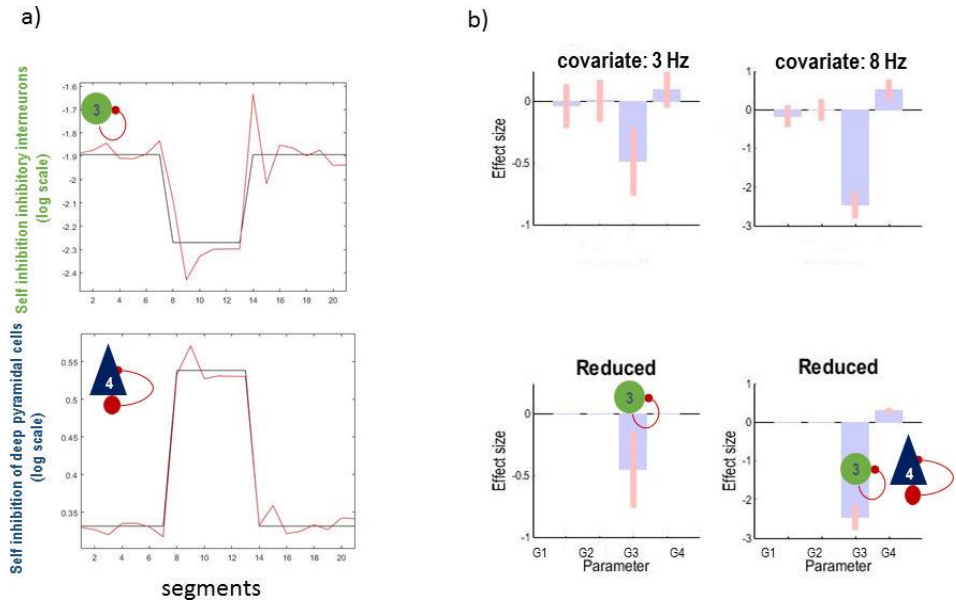

*Figure 8. Parametric empirical Bayes analysis based on the predominant frequency content of the data. a) changes in parameters (black) and predicted changes of G3 and G4 (red) using PEB with three frequency covariates in the GLM design matrix. b) second level GLM parameters (above) and following Bayesian model reduction (below). The BMR suggests that slow waves (3 Hz) are associated with inhibitory populations, whereas higher frequency pathological activity (8 Hz) may be induced by both deep pyramidal and inhibitory populations. $G_1$, $G_2$, $G_3$ and $G_4$ are self-inhibition of excitatory, superficial pyramidal, inhibitory, and deep pyramidal cells, respectively.*

The inferred parameters of the GLM (coupling frequency-specific activity to self-inhibition parameters) are plotted in figure 8. As can been seen from these posterior estimates, the upward/downward changes in self-inhibition of the inhibitory and deep pyramidal cells are well captured. This frequency-specific analysis may provide valuable information that it would not otherwise be possible to extract from electrophysiological data. More importantly, capturing trends in parameter dynamics may be important for designing effective treatments of epilepsy as different intervention mechanisms may have an opposite effect on different ion currents (Blenkinsop et al., 2012, Nevado-Holgado et al., 2012). In the final section, we rehearse P-DCM, using empirical data to provide an illustrative example.

## 4. Worked example: animal model of epilepsy

As the final example in this paper, we apply P-DCM to a Picrotoxin animal model of seizures (Kätzel et al., 2014). In this animal model, the cause of paroxysmal activity can be related to pathological disruption of synaptic connections at the focus of drug injection, which is then spread to other regions (Kätzel et al., 2014, Wood, 2014).

The animal was implanted with a wireless ECoG device on its brain and a cannula was inserted into its primary visual cortex, in the right hemisphere. We then injected a chemo-convulsant drug (Picrotoxin, 300-400nl of 10mM) via the pre-implanted cannula. A few minutes after the injection, large amplitude spikes emerged (due to the hyper-synchronised response of the neuronal population at the site of intervention), followed by occurrence of seizures ~20-30 minutes after injection. Behavioural manifestation of seizures includes freezing, followed by head bobbing (20-30s), hunching, limb-kicking and occasionally rearing, falling over and wet dog shakes. Average duration of the seizures period in this animal model is around ~ 53s, with an inter seizure interval of approximately ~200 s. In this animal model, the activity of the brain recovers to the baseline activity 2-3 hours after the injection, which implies ~15-20 seizures over 2 hrs. A sample of ECoG data and its time frequency response are shown in Figure 9.

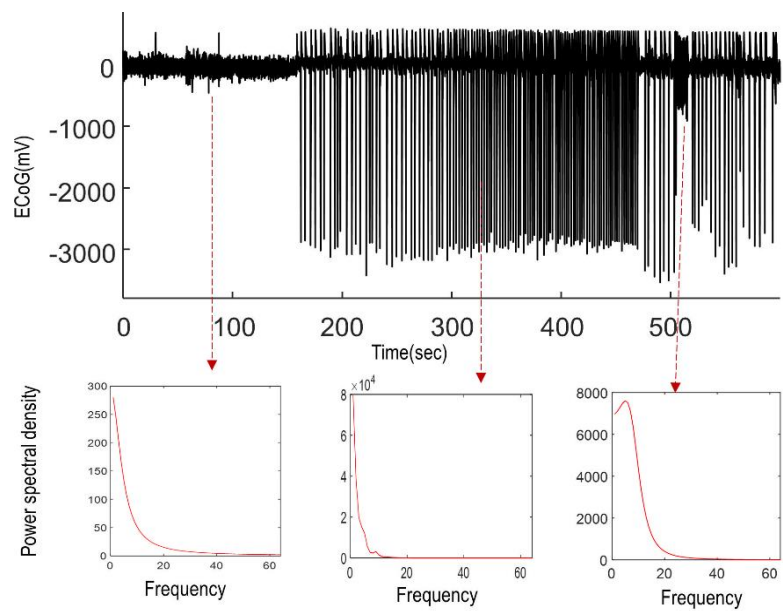

*Figure 9. Real ECoG data from an animal model and spectral responses associated with normal to large spikes and seizures (from left to right respectively).*

At the initial phase of drug injection, seizures commenced focally and only the injection site was affected by the drug. Subsequently other brain regions were recruited by the seizures. We elected to study the underlying causes of the seizures in the initial phase of the pathology and investigate its underlying mechanisms to illustrate an application of P-DCM. See also Appendix B.

We first fitted the CMC model to the normal activity to establish the prior expectation for model parameters. We then epoched the data each of which has 2 seconds duration and estimated key model parameters. We fixed some of the parameters (e.g., noise hyperparameters and sensor gain, which are not likely to vary in this animal model during the experiment) and only allow the rate constant and self-inhibition of neuronal populations to change during and after seizures. After model inversion, we ran a PEB analysis with a binarized spectral envelope as a regressor (column vector with zero and one entries, where zero and one denotes normal and seizure epochs, respectively). The results are shown in Figure 10 and suggest that disinhibition in the inhibitory and deep pyramidal cells populations best explain the data.

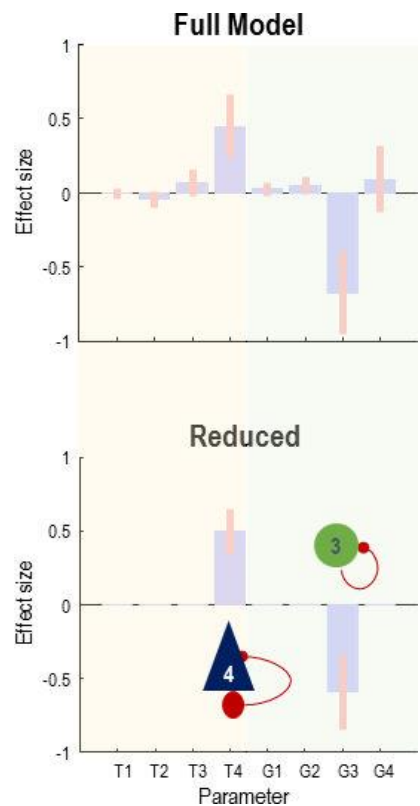

*Figure 10. Parametric empirical Bayes (PEB) and Bayesian model reduction (BMR) of slowly changing parameters. The PEB model is a hierarchical model, with a general linear model (GLM) of the neural parameters at the between-epoch level. The design matrix of the GLM is simply the binarized spectral envelope encoding epochs with and without seizure activity. BMR suggests that disinhibition of deep pyramidal and inhibitory populations was likely to explain the seizure activity. $G_1$, $G_2$, $G_3$ and $G_4$ are self-inhibition parameters of excitatory, superficial pyramidal, inhibitory, and deep pyramidal cells, respectively. $T_1$, $T_2$, $T_3$ and $T_4$ are the time constants of excitatory, superficial pyramidal, inhibitory, and deep pyramidal cells, respectively.*

Our results are in agreement with the clinical understanding of seizures in this animal model (Wood, 2014, Berglind et al., 2014). In detail, these seizures are thought to be due to disruption of interaction between inhibitory populations and other populations (Berglind et al., 2014, Rovainen, 1983). In our modelling results, seizures were explained by reduction of inhibitory self-connectivity (i.e., disinhibition). These results are informative because disruption in the self-regulation of the inhibitory populations engenders changes in postsynaptic interneuron potentials, and thereby affects activity of other populations. Our analysis also suggests that the deep layer is likely to contribute to seizures dynamics, which is plausible, since from an anatomical perspective the density of neurons in the deep layer of the cortex is greater than in the superficial layers. In this DCM, such mechanisms are reflected in the synaptic time constant of deep pyramidal populations that increased by over 50% in this example.

Finally, we repeated the above analysis using frequency-specific regressors to characterise the relationship between disinhibition in particular populations and their frequency specific correlates. Here, the regressors of the GLM were the predominant empirical frequencies in the data of 5, 19, and 40 Hz (identified using singular value decomposition of the time frequency data). The PEB results (Figure 11) show that seizures (which are characterised by 3 to 8 Hz activity) are best explained by the reduction in effective membrane time constant of deep pyramidal cells, with interneurons contributing to 19 Hz activity. This is again consistent with physiological findings from this animal model of seizures.

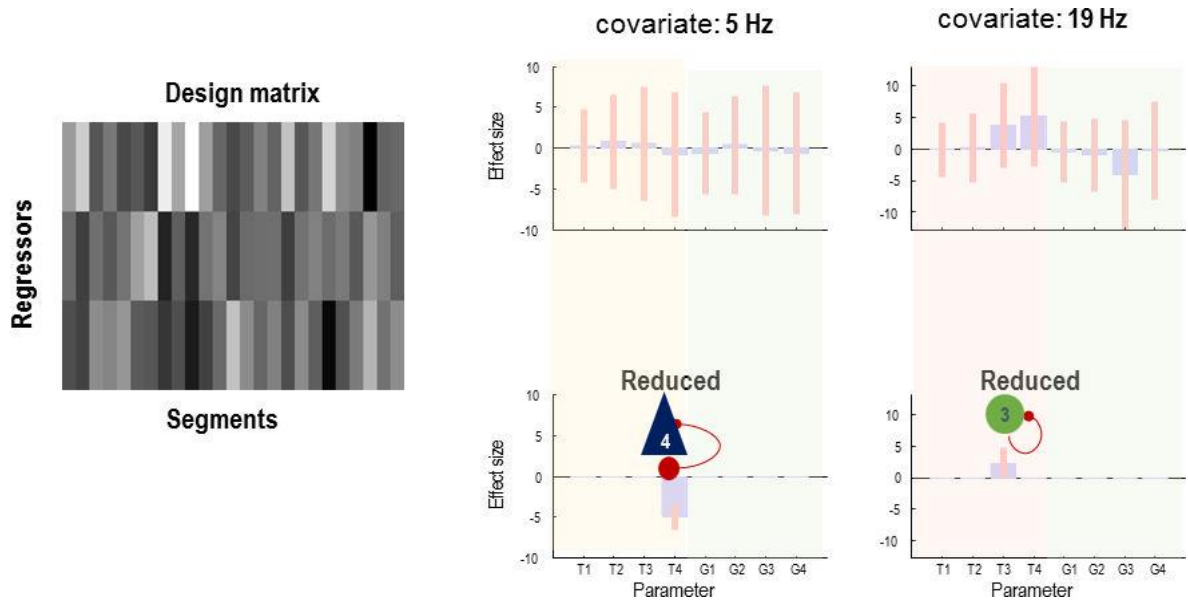

*Figure 11. Parametric empirical Bayes and Bayesian model reduction modelling of slowly changing synaptic parameters. The regressors of the between-epoch GLM comprise the frequency content in the signal (left hand side of the figure). The right side of the figure shows the GLM parameters (top row) and the reduced parameters after Bayesian model reduction (bottom row). The model reduction with this design matrix shows that inhibition and deep pyramidal cells are likely to explain the seizures at the different frequencies. $G_1$, $G_2$, $G_3$ and $G_4$ are self-inhibition parameters of excitatory, superficial pyramidal, inhibitory, and deep pyramidal cells, respectively. $T_1$, $T_2$, $T_3$ and $T_4$ are the time constant of excitatory, superficial pyramidal, inhibitory, and deep pyramidal cells, respectively.*

## 5. Discussion

This paper has introduced a parametric dynamic causal model (P-DCM) that enables one to compare competing hypotheses about the biological mechanisms that might underwrite phase transitions in electrophysiological recordings. Crucially, P-DCM is formulated to elucidate the most likely causal relationships between synaptic efficacy and spectral activity in electrophysiological data, which is commonly used characterise and understand brain states in the healthy and pathological brain (Shaw et al., 2017). The use of an adiabatic approximation and mean field theory furnishes an efficient way to infer the relationship between slow changes in synaptic parameters and their neuronal correlates in spectral data. The resulting model could, in principle, be used to infer how interventions modulate states of the diseased or healthy brain (Fleming et al., 2020, Liang et al., 2015, Moran, 2015). This would require construct validation experiments to test consistency and agreement between predictions of the model and effects of interventions on real brains e.g., (Mina et al., 2013).

We have motivated P-DCM using Synergetic theory and the Adiabatic approximation. As touched on in the introduction, P-DCM can potentially provide complementary information alongside the conventional slow-fast modelling approach. In P-DCM slow variables are modulated with respect to the frequency contents of fast states, the slow variable is equipped with a sort of memory (since the frequency domain representation of neuronal responses rests on the second order statistical moments of fast states). In this respect, one can establish a link between parametric fluctuations in P-DCM and slow dynamics in slow-fast dynamical systems (Sanders et al., 2007). Akin to P-DCM idea, averaging of slow-fast dynamical systems allows separation of slow and fast dynamics, where the slow dynamics are an integral of fast states (under the assumption that fast states are ergodic for each value of the slow states). We also note conceptual links between P-DCM and linear response theory (LRT) in physics (Ruelle, 2009). Specifically, LRT implies that if an ergodic system – e.g., neural mass or mean field model (Marreiros et al., 2009) – is left without any perturbation, it eventually reaches its equilibrium (Lucarini, 2008). If, however, some parameters are slowly changing, LRT assures the existence of a new equilibrium for the system that can be attained instantaneously. In addition, LTR suggests that a process with slowly varying parameters can be reversible (i.e., it can attain its initial equilibrium). Crucially and more closely related to P-DCM, LTR establishes a link between time domain features of dynamical systems and its equilibrium, thereby allowing parametric expression of the system's response in the time/frequency domain (Reick, 2002, Lucarini, 2008, Lucarini et al., 2007, Dykman et al., 1998).

In this work, we offered two forward simulations to illustrate the different kinds of phase transition that one might infer using P-DCM. In the first simulation, phase transitions in brain activity were explained when parameters moved from one region of parameter space to another, i.e. bifurcations (Breakspear et al., 2006). In the second simulation, we show that underlying causes of paroxysmal transitions can be understood as a hysteresis phenomenon (Iasemidis, 2003, Voss et al., 2012), where in one direction of parameter variations, the model produces gamma activity while in reciprocal direction, the model generates pathological spike-wave discharge activity.

In terms of estimating parameters, in the first face validity analysis, we illustrated how one can establish links between fluctuation of parameters and spectral responses in the data. This perhaps would be most useful for tracking dynamics of parameters with respect to changes in spectral responses in real data. The potential application of such a generative model would be tracking, where we would be interested in the modulation of brain activity with respect to changes of transmembrane currents, which may be subject to alteration by interventions. In the second face validity study, we recovered parameter trends that induced spontaneous seizures. In addition, we showed that it is feasible to use Bayesian model reduction to evince underlying biological mechanisms that explained transition dynamics.

In the worked example, we used the CMC model to model the underlying causes of drug-induced seizure activity. Here our results were in line with known pathophysiology. In this example, the underlying causes of seizures were attributed to disruption of inhibition. Our modelling approach suggests that the inhibitory population is largely responsible for the generation of 3 Hz waveforms. While our results clearly need to be further reproduced and validated; they speak to a promising application of P-DCM. As measured above, we only considered the initial phase of seizures in this animal model, where underlying causes of the seizures can be linked with the local effects of the drug. More interesting questions might be investigated using P-DCM in the second phase of seizures in this animal model, where other brain regions are contributing to the pathology.

A key theme in this modelling endeavour is the circular causality between connection strengths or synaptic efficacy and the neuronal activity these connections support. There is a large literature on models of synaptic plasticity (Demšar and Forsyth, 2020) and associative plasticity (Humeau et al., 2003, Kujirai et al., 2006); namely, the relationship between pre-and post-synaptic activity that can be linked to slow variation of synaptic parameters. One could also consider activity-dependent plasticity (Rebola et al., 2010, Isomura and Friston, 2019) as a function of the complex cross spectra or cross covariance functions summarising neuronal dynamics. In turn, this means that one could appeal to the notion of spike timing-dependent plasticity (STDP), to model changes in effective connectivity in terms of the cross-covariance function between a source and target population (which is well formulated in the spectral domain). The utility of expressing plasticity (i.e., the dynamics of slow parameters) in terms of covariance functions is that there is an equivalent frequency space representation; thereby accommodating the representation of dynamics in terms of complex cross spectra in the generative (dynamic causal) model. The only unknown hyperparameters in this instance are the coefficients scaling the amplitude and width of the functions that lead to increases and decreases in synaptic efficacy. Furthermore, there are empirical constraints on these functions. For example, for STDP, one would normally use the first derivative of a Gaussian function, with a dispersion of about 50 ms. With this formulation of parametric dynamics, we automatically account for ideas like spike rate adaptation (Peron and Gabbiani, 2009), spike timing dependent plasticity (Caporale and Dan, 2008, Dan and Poo, 2006) and the transmission delays inherent in the dynamic causal modelling of cross spectra (Friston et al., 2012). In the future, we hope to extend P-DCM to address the delicate interplay between synaptic plasticity and spectral fluctuations.

## 6. Appendix A

A technical aspect of this generative model is the form of the likelihood for the complex cross spectra. With ideal estimators, one could assume that these spectral data features had a Wishart distribution, with one degree of freedom for each frequency. However, we can assume that the cross spectra constitute the average of estimates, with consistent and asymptotically normal estimates. In this setting, the variance of the difference between the predicted and observed spectral estimates is equal to the cross spectral density times the (effective) degrees of freedom ($h$), which we treat as an unknown parameter. Specifically, the precision is given by the asymptotic results where, the scaled difference between the sample spectral density ($g$) and the predicted density ($G$) (Camba-Mendez and Kapetanios, 2005):

$$e = vec\,(g - G) \qquad (A.1)$$

is asymptotically complex normal and the covariance between $e(i,j)$ and $e(u,v)$ is given by Q/h and:

$$Q = G(i,u) * G(j,u); \; h = 2 \times m + 1$$

Here *m* represents the number of averages from a very long time series. The inverse of the covariance is thus a scaled precision, where the hyperparameter ($h$) plays the role of the degrees of freedom (e.g., the number of averages comprising the estimate). We use the sample spectral density to create a frequency specific precision matrix for the vectorised spectral densities, under the assumption that the form of this sample spectral density resembles the predicted spectral density (which will become increasingly plausible with convergence).

# 6. Appendix B

In the animal model of seizures in this paper, paroxysmal activity may be induced as a result of the disruption of the interaction between inhibitory interneurons with other populations. Such disruption can be modelled by changing the balance between synaptic efficacies and gains of excitatory and inhibitory populations (Wendling et al., 2002). For example, since the drug was delivered to the deep layer of cortex, one may alter the rate constant/self-inhibition of inhibitory, spiny stellate excitatory cells and deep pyramidal cells in the CMC model to replicate both spike discharges and recurrent of seizures. One example of such a forward simulation is shown in Figure 12.

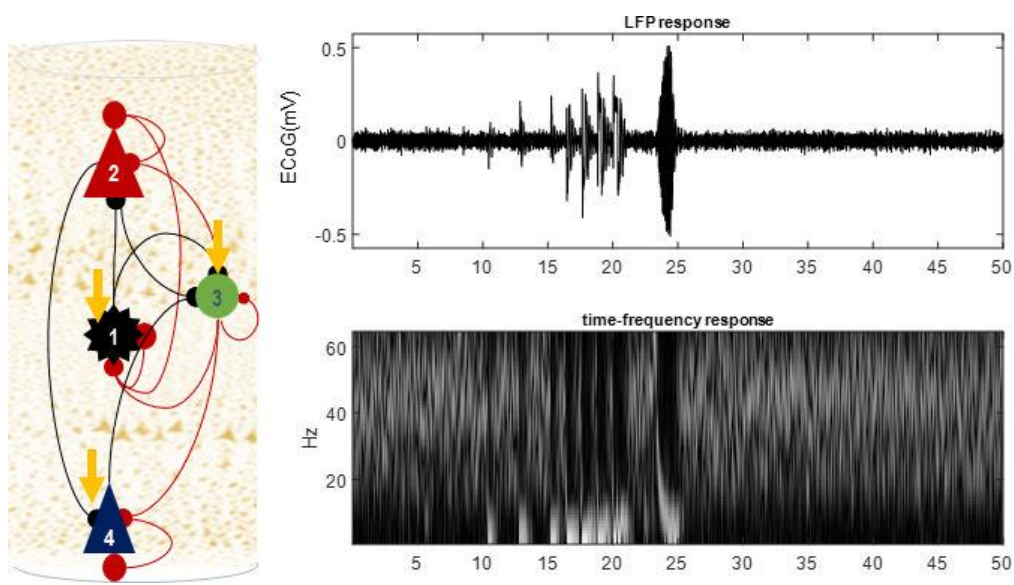

*Figure 12. Forward simulation of the CMC model with time varying parameters. Here the rate constants and synaptic efficacy of inhibitory, spiny stellate and deep pyramidal cells are changed from their prior expectation to generate spike-like activity and recurrent of seizures, as shown in LFP response of model.*

One should note that, this is only one way to generate pathophysiology and there may be many other plausible mechanisms that one could implement using the CMC model to generate paroxysmal transitions of the data. As explained in the paper, the most compelling questions – for epileptologists – are which synaptic parameters are likely to explain the underlying causes of pathognomonic data?

# 7. Acknowledgments

The Wellcome Centre for Human Neuroimaging is supported by core funding from Wellcome [203147/Z/16/Z]. RW holds a Senior Research Fellowship funded by the Worshipful Company of Pewterers.

Table 1: Parameters of the neuronal model (see also Figure 2)

|   | Description | Parameterisation | Prior |
|---|---|---|---|
| T | Postsynaptic rate constant | $exp(\theta_\kappa) \cdot T_i$ <br> $T = [256, 128, 16, 32]$ | $p(\theta_\kappa) = N(0, 1/16)$ |
| g | Intrinsic connectivity between populations *i* and *k* in each region | $exp(\theta_a) \cdot g$ | $p(\theta_a) = N(0, 1/16)$ |
| G | Self-inhibitory connection | $exp(\theta_a) \cdot G$ | $p(\theta_a) = N(0, 1/16)$ |

Table 2: Glossary of variables and expressions

| Variable | Description |
|---|---|
| $u$ | Exogenous input |
| $v$ | The *i*-th (neuronal) state in region *j*; e.g., mean depolarisation of a neuronal population |
| $s(v)$ | The neuronal firing rate – a sigmoid squashing function of depolarisation |
| $L$ | Lead field vector mapping from (neuronal) states to measured (electrophysiological) responses |
| $g_v(\omega), g_o(\omega), g_y(\omega)$ | Spectral density of (neuronal) state fluctuations, observation error and ensuing measurement, respectively |
| $\partial_x f$ | System Jacobian or derivative of system flow with respect to (neuronal) states |
| $k(t) = FT[K(\omega)]$ | First order kernel mapping from inputs to responses; c.f., an impulse response function of time. This is the Fourier transform of the transfer function |
| $K(\omega) = FT[k(t)]$ | Transfer function of frequency, modulating the power of endogenous neuronal fluctuations to produce a (cross spectral density) response. This is the Fourier transform of the kernel |